\documentstyle[psfig]{mn}

\newif\ifAMStwofonts

\ifoldfss
  \ifCUPmtlplainloaded \else
    \NewTextAlphabet{textbfit} {cmbxti10} {}
    \NewTextAlphabet{textbfss} {cmssbx10} {}
    \NewMathAlphabet{mathbfit} {cmbxti10} {} 
    \NewMathAlphabet{mathbfss} {cmssbx10} {} 
  \fi
  \ifAMStwofonts
    \ifCUPmtlplainloaded \else
      \NewSymbolFont{upmath} {eurm10}
      \NewSymbolFont{AMSa} {msam10}
      \NewMathSymbol{\upi}     {0}{upmath}{19}
      \NewMathSymbol{\umu}     {0}{upmath}{16}
      \NewMathSymbol{\upartial}{0}{upmath}{40}
      \NewMathSymbol{\leqslant}{3}{AMSa}{36}
      \NewMathSymbol{\geqslant}{3}{AMSa}{3E}

      \let\leq=\leqslant 
      \let\geq=\geqslant 
    \fi
  \fi
\fi 

\ifnfssone
  \newmathalphabet{\mathit}
  \addtoversion{normal}{\mathit}{cmr}{m}{it}
  \addtoversion{bold}{\mathit}{cmr}{bx}{it}
  \newmathalphabet{\mathbfit} 
  \addtoversion{normal}{\mathbfit}{cmr}{bx}{it}
  \addtoversion{bold}{\mathbfit}{cmr}{bx}{it}
  \newmathalphabet{\mathbfss} 
  \addtoversion{normal}{\mathbfss}{cmss}{bx}{n}
  \addtoversion{bold}{\mathbfss}{cmss}{bx}{n}
  \ifAMStwofonts
    \ifCUPmtlplainloaded \else
      \UseAMStwoboldmath
      \makeatletter
      \new@mathgroup\upmath@group
      \define@mathgroup\mv@normal\upmath@group{eur}{m}{n}
      \define@mathgroup\mv@bold\upmath@group{eur}{b}{n}
      \edef\UPM{\hexnumber\upmath@group}
      \new@mathgroup\amsa@group
      \define@mathgroup\mv@normal\amsa@group{msa}{m}{n}
      \define@mathgroup\mv@bold\amsa@group{msa}{m}{n}
      \edef\AMSa{\hexnumber\amsa@group}
      \makeatother
      \mathchardef\upi="0\UPM19
      \mathchardef\umu="0\UPM16
      \mathchardef\upartial="0\UPM40
      \mathchardef\leqslant="3\AMSa36
      \mathchardef\geqslant="3\AMSa3E

      \let\leq=\leqslant 
      \let\geq=\geqslant 
    \fi
  \fi
\fi 

\ifnfsstwo
  \DeclareMathAlphabet{\mathbfit}{OT1}{cmr}{bx}{it}
  \SetMathAlphabet\mathbfit{bold}{OT1}{cmr}{bx}{it}
  \DeclareMathAlphabet{\mathbfss}{OT1}{cmss}{bx}{n}
  \SetMathAlphabet\mathbfss{bold}{OT1}{cmss}{bx}{n}
  \ifAMStwofonts
    \ifCUPmtlplainloaded \else
      \DeclareSymbolFont{UPM}{U}{eur}{m}{n}
      \SetSymbolFont{UPM}{bold}{U}{eur}{b}{n}
      \DeclareSymbolFont{AMSa}{U}{msa}{m}{n}
      \DeclareMathSymbol{\upi}{0}{UPM}{"19}
      \DeclareMathSymbol{\umu}{0}{UPM}{"16}
      \DeclareMathSymbol{\upartial}{0}{UPM}{"40}
      \DeclareMathSymbol{\leqslant}{3}{AMSa}{"36}
      \DeclareMathSymbol{\geqslant}{3}{AMSa}{"3E}

      \let\leq=\leqslant 
      \let\geq=\geqslant 
    \fi
  \fi
\fi 

\ifCUPmtlplainloaded \else
  \ifAMStwofonts \else 
    \def\upi{\pi}
    \def\umu{\mu}
    \def\upartial{\partial}
  \fi
\fi

\title{Constraining the cosmic equation of state from old galaxies at high
redshift}

 \author[J. A. S. Lima and J. S. Alcaniz]
        {J. A. S. Lima and J. S. Alcaniz \\
         Universidade Federal do Rio Grande do Norte,
 Departamento de F\'{\i}sica Te\'orica e Experimental,\\
Caixa Postal 1641, 59072-970 Natal, RN, Brasil}

\pagerange{\pageref{firstpage}--\pageref{lastpage}}
\pubyear{1999}

\begin{document}

\maketitle

\label{firstpage}

\begin{abstract}
New limits on the cosmic equation of state are derived from age measurements
of three recently reported old high redshift galaxies (OHRG). The results are
based on a flat FRW type cosmological model driven by nonrelativistic matter
plus a smooth component parametrized by its equation of state $p_{x} =
\omega\rho_{x}$ ($\omega \geq -1$). The range of $\omega$ is strongly
dependent on the matter density parameter. For $\Omega_{M} \sim 0.3$, 
as indicated from dynamical measurements, the age estimates of the OHRG
restricts the cosmic parameter to $\omega\leq - 0.27$. However, if
$\Omega_{M}$ is the one suggested by some studies of field galaxies, i.e,
$\Omega_{M} \simeq 0.5$, only a cosmological constant ($\omega=-1$) may be
compatible with these data.  
\end{abstract}


Recent distance measurements of some Type Ia supernovae at intermediary and high redshifts 
indicate that the expansion of the Universe is speeding up, rather than
slowing down (Riess et al. 1998; Perlmutter et al. 1999a). Indirectly, these 
results also mean that the Universe is much older than the one  predicted by
the standard CDM flat model with a critical deceleration   parameter  ($q_{o}
=$ 0.5). If confirmed from more accurate observations and/or a different class
of phenomena, the existence of these data poses a crucial problem for all CDM
models since their generic prediction is a decelerating universe ($q_{o} >
0$), whatever the sign adopted for the curvature parameter. 

Another source of difficulties for the standard CDM model is related to the ``age problem" or its 
modern variant, the age of old high redshift objects. It should be
recalled  that some measurements of the Hubble parameter yielded  $h = H_{o}/100
\rm{km/sec/Mpc} = 0.7 \pm 0.1$ (Nevalainen \&
Roos 1998, Friedmann 1998). In particular, this means that the expansion age for a FRW
flat matter dominated universe ($t_o = {2 \over 3}H_o^{-1}$) falls 
within the interval  $8.3 \rm{Gyr} \leq  t_o \leq 10.5 \rm{Gyr}$, while the
age inferred from globular clusters lies typically into the range $t_{gc} \sim
13-15\rm{Gyr}$ or higher (Bolton \& Hogan 1995; Pont et al. 1998). As widely
known, this conflict is not alliviated if SNe data are considered. In the age
analysis of Perlmutter at al. (1998), the favored value is $h=0.63$, while
Riess et al. (1998) found $h=0.65 \pm 0.02$, from their SNe data.

Age measurements of extragalactic
objects at high redshifts also provide an alternative route to the ``age
problem". In this case, the discoveries of a $4.0-\rm{Gyr}$-old galaxy at
$z=1.175$ (Stockton et al. 1995), of a $3.5-\rm {Gyr}$-old galaxy at $z=1.55$
(Dunlop et al. 1996; Spinrard et al. 1997), and of a $4.0-\rm{Gyr}$-old galaxy at $z=1.43$ 
(Dunlop 1998) have been   proved to be incompatible with age estimates for a
flat Universe unless the Hubble parameter is very low. These constraints are
even more stringent than the ones from globular cluster age measurements
(Dunlop 1996, Krauss 1997, Roos \& Harun-or-Raschid 1998). More recently, it
was shown that the existence of the two OHRG discovered by Dunlop would be
accommodated in a model with no cosmological constant only if  $\Omega_M \leq
0.37$ (Alcaniz and Lima 1999).

In the last few years, flat models with a relic cosmological constant ($\Lambda$CDM) have also 
been considered as a serious candidate for standard cosmology (Krauss \&
Turner 1995; Krauss 1997). However, although fitting some observations better
than other theoretical models (e.g., the first accoustic peak of the relic
radiation angular power spectrum), $\Lambda$CDM cosmologies are reasonably
restricted by the statistics of gravitational lenses (SGL) (Kochanek 1996,
Falco et al. 1998, Waga and Miceli 1999). On the other hand, though the method
based on OHRG have given a lower limit of $\Omega_\Lambda \geq 0.5$ (Alcaniz
and Lima 1999), from a theoretical viewpoint, these models are plagued by a
profound contradiction: in order to dominate the dynamics of the Universe only
at recent times, a very small value for the cosmological constant
($\Lambda_{o}  \sim 10^{-56}\rm{cm}^{-2}$) is required from observations,
while naive estimates based on quantum field theories are 50-120 orders of
magnitude larger, thereby originating an extreme fine tunning problem
(Weinberg 1989, Sahni and Starobinsky 1999). 

Cosmologies containing an extra component describing the dark matter, and simultaneously 
accounting for the present accelerated stage of the universe have also been
widely discussed in the literature. Indeed, the absence of a convincing
evidence on the nature of the dark component has stimulated the debate and
theoretical speculations.  Some possible candidates are: a time varying
$\Lambda$-term (Ozer \& Taha 1986, 1987; Freese et al. 1987; Carvalho et al.
1992; Waga 1993; Lima and Maia 1994; Lima and Trodden 1996; Lima 1996; 
Silveira and Waga 1997), a relic scalar field (Peebles 1984; Ratra and Peebles
1988; Caldwell et al. 1998; Maia and Lima 1999; Lima et al. 2000). Sometimes, the extra
component is named ``X-matter", or ``quintessence", which is simply
characterized by an arbitrary equation of state $p_x=\omega\rho_{x}$, where
$\omega\geq -1$ (Turner and White 1997; Chiba et al. 1997; Efsthatiou 1999;
Lima and Alcaniz 2000). In this case, constraints from large scale structure
(LSS) and cosmic microwave background anisotropies (CMB) complemented by the
SN Ia data, require $0.6 \leq \Omega_x \leq 0.7$ and $\omega < -0.6$ ($95\%$
C.L.) for a flat universe (Perlmutter et al. 1999b; Efsthatiou 1999), while for
universes with arbitrary spatial curvature the limit is $\omega < -0.4$
(Efstathiou 1999).

In the present work, we focus our attention to this kind of
``quintessence" or ``X-matter" cosmology. As a matter of fact, due to their generality these
models merit a broader discussion. In principle, 
to check the validity of a theory or model (for instance, the $\Lambda$CDM
model), it is interesting to insert it in a more general framework, herein
quantified by the $\omega$ parameter. Taking the limiting case $\omega = -1$,
the $\Lambda$CDM results are readily recovered. 

In this context, by considering the three above mentioned OHRG, we derive new limits on the
quintessence parameter $\omega$. In particular, by extending  the method
proposed in a previous paper (Alcaniz and Lima 1999), we show that $\omega
\leq-0.2$ and $\omega
\leq-0.4$ if the density parameter  lies in the observed range
$\Omega_{M} \sim 0.2 - 0.4$ (Dekel et al. 1996), with the lower value of $\Omega_M$ corresponding to
higher $\omega$.  

For a spatially flat, homogeneous,
and isotropic cosmologies driven by nonrelativistic matter and a separately
conserved exotic fluid with equation of state, $p_{x} = \omega\rho_{x}$,
the Einstein field equations can be written as: 
\begin{equation} 
({\dot{R} \over R})^{2} = H_{o}^{2}\left[\Omega_{M}({R_{o} \over R})^{3} +  
\Omega_x({R_{o} \over R})^{3(1 + \omega)}\right] \quad , 
\end{equation} 
\begin{equation} 
{\ddot{R} \over R} = -{1 \over 2}H_{o}^{2}\left[\Omega_{M}({R_{o} \over  
R})^{3} +  
(3\omega + 1)\Omega_x({R_{o} \over R})^{3(1 + \omega)}\right]  
\quad , 
\end{equation} 
where an overdot denotes derivative with respect to time, $H_{o}$ is the  
present value of the Hubble parameter, and $\Omega_{M}$ and $\Omega_x$ 
are the present day matter  and quintessence density parameters. As one may
check  from (1) and (2), the case $\omega= - 1$ corresponds effectively to a
cosmological constant. The age-redshift relation for FRW type universes with
this extra smooth component reads
\begin{eqnarray} 
t(z) & = &H_{o}^{-1}\int_{o}^{(1 + z)^{-1}}{dx \over x\sqrt{\Omega_{M}x^{-3}
+   \Omega_{x} x^{-3(1 + \omega)}}} \nonumber \\ 
&   &= H_{o}^{-1}f(\Omega_{M},
\omega, z)    \quad , 
\end{eqnarray}
where the flat condition constraint, $\Omega_{M} = 1 -  \Omega_x$, has been
inserted.    

Before proceed further, we call attention for an important
point: for a fixed value of the  density parameter $\Omega_M$, the age  
of the Universe predicted by this ``quintessence" model decreases with the increasing of 
$\omega$. Hence, taking for granted that the age of the Universe in a
given redshift is bigger than or at least equal to the age of its oldest
objects, the  existence of these OHRG give rise to an upper bound for
$\omega$. Let us now introduce the
dimensionless ratio 

\begin{equation} 
{t(z) \over t_{g}} = {f(\Omega_{M}, \omega, z) \over H_{o}t_{g}} \geq 1  
\quad, 
\end{equation} 
where $t_{g}$ is the age of an arbitrary object, and $f(\Omega_{M}, \omega, z)$ is the 
dimensionless integral factor appearing in the expression for t(z). For each
extragalactic object, this inequality defines a dimensionless parameter
$T_g=H_{o}t_{g}$. In particular, for the LBDS 53W091 radio galaxy discovered by
Dunlop et al. (1996), the lower limit to the age of this galaxy yields
$T_{G}(1.55) = 3.5H_o\rm{Gyr}$, which take values on the interval  $0.21\leq
T_G \leq 0.28$.  The extreme values of $T_G$ have been determined by the error
bar of $h$. It  thus follows that $T_G \geq 0.21$, and from (2) we see that at
this $z$ the matter dominated flat FRW model furnishes an age parameter
$T_{FRW} \leq 0.16$, which is far less than the previous value of $T_G$. 
Naturally, for a given value of $h$, only models having an expanding age
parameter bigger than the corresponding value of $T_G$ at $z=1.55$ will be
compatible with the existence of this galaxy. The standard Einstein-de Sitter
FRW model is (beyond doubt) ruled out  by this test (Alcaniz and Lima 1999). 

\begin{figure}   
\centerline{\psfig{figure=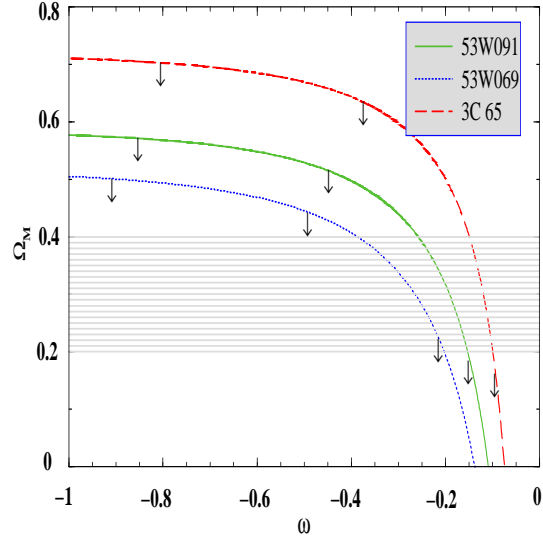,width=3truein,height=3truein}  
\hskip 0.1in} 
\caption{Contourns of fixed age parameters $T_g = H_{o}t_{g}$
for the three OHRG  reported above. Solid curve correspond to LBDS 53W091,
dashed lines to the radio galaxy 3C 65, and dotted lines to the LBDS radio
galaxy 53W069.  As explained in the text, the contours are obtained for the
minimal values of $T_g$. For each contour the arrows point to the allowed
parameter space, while the shadowed horizontal region yields the observed
interval of $\Omega_M$. We see that  the more restrictive upper limit is
provided by the radio galaxy 53W069 (see also table 1).}    
\end{figure}

In order to assure the robustness of the limits, two conditions have  sistematically been adopted in 
our computations: 

(i) The minimal value for the Hubble parameter. In this case, we use the one
obtained by the HST Key project, i.e., the  round number value $H_{o} = 60
\rm{km/sec/Mpc}$ (Friedmann 1998).  
 
ii) The underestimated age for all old high redshift galaxy. 

The above conditions are almost self-explanatory when interpreted in the spirit of inequality 
(4). First, the smaller the  value of $H_{o}$, the larger the age predicted
by the model, and,  second, objects with smaller ages are more easily
accommodated, thereby guaranteeing that the model is always favored in the
estimates presented here. Indeed, concerning the value of $h$, and specially
its lower bound, we are being rather conservative since it was recently
updated to nearly $10\%$ of accuracy ($h = 0.71
\pm 0.07$, $1\sigma$) by Friedman and collaborators (1999), and the data from
SNe also point consistently to $h > 0.6$ or even higher (Perlmutter et al. 1998, Riess et al. 1998). 
On the other hand, we  also recall that the
best-fitting spectral synthesis models has indicated strong evidence for a
minimum age of 4.0 Gyr for the 3C 65 ($z = 1.175$) (Stockton et al. 1995), of
3.5 Gyr for the LBDS 53W091 ($z = 1.55$), and 4.0 Gyr for the LBDS 53W069 ($z
= 1.43$) (Dunlop et al. 1996; Spinrad et al. 1997; Dunlop  1998; Dunlop 1999).
Even taking into account the above conditions, the  discrepancy  between these
 observational values and the predictions of a flat matter dominated model is
evident. For instance, if $h = 0.6$, the age predicted by this model for an
object at $z = 1.175$ is $t_{z} \leq$ 3.35 Gyr, while for an object at $z =
1.55$ is  $t_{z} \leq$ 2.66 Gyr. For a  flat universe  with cosmological
constant ($\omega= -1$), these data may be fitted only if the vacuum energy
contribution is $\Omega_{\Lambda} \geq 0.29$ and $\Omega_{\Lambda} \geq 0.42$,
 respectively. The situation is even worse if one considers the object at $z =
1.43$ with a minimal age of 4.0 Gyr (LBDS 53W069). In this  case, the age
predicted is $t_{z} \leq$ 2.85 Gyr and the vacuum energy contribution  should
be $\Omega_{\Lambda} \geq 0.5$ (Alcaniz \& Lima 1999). 

\begin{table*} 
\centering 
\begin{minipage}{140mm} 
\caption{Limits to $\omega$ for a given $\Omega_{\rm{M}}$} 
\begin{tabular}{rlll} 
\hline \hline
\\ 
\multicolumn{1}{c}{Method}& 
\multicolumn{1}{c}{Reference}& 
\multicolumn{1}{c}{$\Omega_{\rm{M}}$ (flat)}& 
\multicolumn{1}{c}{$\omega$}\\ 
\\ 
\hline \hline 
\\ 
CMB + SNe Ia:........& Turner \& White (1997)& $\simeq 0.3$& $\simeq
-0.6$\\   
 & Efstathiou (1999)& $\sim$&  $< -0.6$\\ 
SNe Ia......................& Garnavich et al. (1998)&  
$\sim$& $< -0.55$\\ 
SGL + SNe Ia..........& Waga \& Miceli (1999)& 
$0.24$& $< -0.7$\\ 
SNe Ia + LSS..........& Perlmutter et al. (1999)& $\sim$ & $< -0.6$\\ 
Various.....................& Wang et al. (2000)& $0.2 -0.5$& $ < -0.6$\\ 
Old High-z Galaxies: & & \\
$z = 1.43$.................& This paper&0.3& $< -0.27$\\
$z = 1.43$.................& This paper& $\geq 0.5$& $-1$\\
$z = 1.55$.................& This paper& 0.3& $<-0.20$\\
$z = 1.55$.................& This paper& 0.5& $<-0.40$\\
$z = 1.175$...............& This paper& 0.3& $<-0.12$\\
$z = 1.175$...............& This paper& 0.5& $<-0.20$\\
\hline 
\hline 
\end{tabular} 
\end{minipage} 
\end{table*} 

In Fig. 1, we display the parameter space $\Omega_{M} - \omega$. For a given OHRG, each 
contour represent the minimal value of its age parameter ($T_g = H_{o}t_{g}$)
in the respective redshift. If this parameter is greather, the curves are
displaced as suggested by the arrows in the picture, that is, for the inner
region of each contourn. Thus, if $T_g$ increases the available parameter
space is diminished, or equivalently, for a given redshift z, older galaxies
require smaller values of the pair ($\Omega_M, \omega$). The shadowed
horizontal region corresponds to the observed range $\Omega_m = 0.2-0.4$
(Dekel et al. 1996), which is used to fix the upper limit to the cosmic parameter. 
Note that the allowed range for $\omega$ is reasonably large. For example, if
$\Omega_{M} \sim 0.3$, as sugested by dynamical estimates on scale up to about
2$h^{-1}$ Mpc (Calberg et al.1996; Bahcall \& fan 1998), the age-redshift
relation for the LBDS 53W091 contrains $\omega$ to be $\leq -0.20$. If
$\Omega_M$ is the one derived by some analyses of large-scale structure and
field galaxies, i.e., $\Omega_{M} \sim 0.5$ (Tammann 1998), we find $\omega
\leq -0.4$.  For the radio galaxy 3C 65 at $z = 1.175$, the corresponding
interval $\Omega_M = 0.3-0.5$ provides $\omega \leq -0.12$ and $\omega \leq
-0.20$, respectively.  The most restrictive upper bounds on  $\omega$ comes
from the radio galaxy LBDS 53W069. In this case, for $\Omega_{M} \sim 0.3$, 
we have $\omega \leq -0.27$ whereas for  $\Omega_{M} \sim 0.5$, only a
$\Lambda$CDM model ($\omega=-1$) is  compatible with the minimal value of its
age parameter $T_g$. In particular these results  agree with the 1$\sigma$
upper limit derived by Waga and Miceli (1999) using  statistics of strong
gravitational lenses (SGL) and high-z type Ia supernovae ($\omega <   -0.7$),
as well as with the  2$\sigma$ upper limit obtained  by Efstathiou (1999) and
Perlmutter et. al (1999b) using high-z type Ia supernovae and cosmic microwave 
background anisotropies ($\omega < -0.6$). As one may see from Fig. 1, for
$\omega = -1$ (cosmological constant) the results above mentioned are
recovered (for more details see Alcaniz \& Lima 1999).

At this point it is interesting to compare our results with some recent determinations of $\omega$ 
derived from independent methods.
Recently, Garnavich et al. (1998) using the SNe Ia data from the High-Z
Supernova Search Team (Riess et al. 1998) found $\omega < -0.55$ ($95\%$ C.L.) for flat models 
whatever the value of $\Omega_M$ whereas for arbitrary
geometry they obtained $\omega < -0.6$ ($95\%$ C.L.). As commented there,
these values are inconsistent with a unknown component like topological
defects (domain walls, string, and textures) whose $\omega = - \frac{n}{3}$,
being $n$ the dimension of the defect. The results by Garnavich et al. (1998)
agree with the constraints obtained from a wide variety of different phenomena 
(Wang et al. 2000), using the ``concordance cosmic" method. Their combined
maximum likelihood analysis suggests $\omega \leq -0.6$, which is more
stringent than the upper limits derived here, unless the density parameter is
slightly larger than the observed range ($\Omega_M = 0.2-0.4$). The main
results of the present paper together with other determinations of $\omega$
are summarized in Table 1. 

Finally, we stress that the new constraints on the ``quintessence" 
parameter presented here reinforce the importance of old high redshift
galaxies as special probes to the late stages of the universe. Even taking a
too conservative viewpoint that such constraints are only suggestive (perhaps
due to an unknown systematic effect on the data), our results point
consistently to the same direction, namely: if $\Omega_M > 0.4$ a cosmological
constant ($w=-1$) is favored by the existence of OHGRs (see Table 1). This
conclusion is also supported by a more detailed analysis combining the age of
the universe problem and the $H_{o}-\omega$ diagram. Thus, it should be
interesting to insert this high redshift method and the related constraints
within the large set of quintessence cosmological tests recently discussed by
Wang et al. (2000).

\section*{Acknowledgments}

This work was partially 
supported by the project Pronex/FINEP (No. 41.96.0908.00) and 
Conselho Nacional de Desenvolvimento Cient\'{\i}fico e 
Tecnol\'ogico - CNPq (Brazilian Research Agency).

\bsp

\label{lastpage}

\end{document}